\documentstyle[12pt]{article}
\def\greaterthansquiggle{\raise.3ex\hbox{$>$\kern-.75em\lower1ex\hbox{$\sim$}}}
\def\lessthansquiggle{\raise.3ex\hbox{$<$\kern-.75em\lower1ex\hbox{$\sim$}}}
\newcommand{\beq}{\begin{equation}}
\newcommand{\eeq}{\end{equation}}
\newcommand{\beqa}{\begin{eqnarray}}
\newcommand{\eeqa}{\end{eqnarray}}

\newcommand{\grts}{\greaterthansquiggle}
\newcommand{\lets}{\lessthansquiggle}

\newcommand{\ra}{\rightarrow}

\newcommand{\vp}{\varphi}
\newcommand{\vt}{\vartheta}

\newtheorem{theorem}{Theorem}
\newtheorem{lemma}{Lemma}
\newtheorem{prp}{Proposition}
\newtheorem{cor}{Corollary}

\def\au{{\setbox0=\hbox{\lower1.36775ex%
\hbox{''}\kern-.05em}\dp0=.36775ex\hskip0pt\box0}}
\def\ao{{}\kern-.10em\hbox{``}}

\normalsize
\begin{document}
\bibliographystyle{plain}

\begin{titlepage}
\begin{flushright}
ESI preprint 120\\
July 1994
\end{flushright}
\vfill
\begin{center}
{\Large \bf
Harmonic maps between three-spheres} \\[40pt]
Piotr Bizo\'{n}\\
Erwin Schr\"odinger Institut f\"ur Mathematische Physik \\
 Wien, Austria \\
and \\
Institute of Physics, Jagellonian University \\
Cracow, Poland\\
e-mail: bizon@ztc386a.if.uj.edu.pl
\vfill
{\bf Abstract}
\end{center}

It is shown that
smooth maps $f: S^3 \rightarrow S^3$ contain two countable families
of harmonic representatives in the homotopy classes of degree zero
and one.
\vfill
\end{titlepage}

\section{Introduction}
Let $f:M \ra N$ be a smooth map between Riemannian manifolds
$(M, g)$ and $(N, h)$.
The energy of the map $f$ is defined by
\beq
E(f) = \int_{M} h_{AB}(f) \, \frac{\partial f^A}{\partial x^i}
\frac{\partial f^B}{\partial x^j} \, g^{ij} \, dV_M \, ,
\eeq
where $x^i$ denote local coordinates in $M$ and $f^A$ denote local
coordinates of the point $f(x)$ in $N$. The critical points of
this energy functional are called harmonic maps [1] (in physical literature
the model defined by (1) is called the $\sigma$-model).

In this paper I consider the case when both $M$ and $N$ are
three dimensional spheres $S^3$.
Let $(\psi,\vt,\vp)$ be standard hyperspherical coordinates
on $M$
 in which $ds^2=d\psi^2 + \sin^2 \!\psi \,(d\vt^2
+ \sin^2 \!\vt\, d\vp^2)$, and let $(\Psi,\Theta,\Phi)$ be similar
coordinates on the target space $N$. Assume further that the map
$f$ is $SO(3)$ symmetric:
\beq
 \Psi = f(\psi), \qquad  \Theta=\vt, \qquad \Phi = \vp \, .
\eeq
Then the energy functional (1) reduces to
\beq
E(f) = 4 \pi \int_0^{\pi} \left( f'^2 + 2 \, \frac{\sin^2\!f}
{\sin^2 \!\psi} \right)\,\sin^2 \!\psi \, d\psi \, ,
\eeq
and the corresponding Euler-Lagrange equation is
\beq
(\sin^2 \!\psi \, f')' -\sin(2f) = 0.
\eeq
By the principle of symmetric criticality [2] the solutions of this equation
 are
the critical points of the energy functional (1), and therefore
determine harmonic maps.
 The reduced model is invariant under the discrete transformations
$f \ra f+n\pi$ and $f \ra -f$. To get partially rid of this degeneracy I
shall work in the configuration space $\cal C$ of equivalence classes defined
by
$f \sim \tilde{f}$ iff $\cos f = \cos \tilde{f}$. Note that in $\cal C$
 the two functions $f$ and $\bar{f} \equiv \pi-f$ are
 considered as distinct.

I am interested in  solutions of Eq.(4)
for which the energy density
\beq
\rho =  f'^2 + 2 \, \frac{\sin^2\!f}{\sin^2 \!\psi}
\eeq
is everywhere finite. Such solutions will be referred to as regular.
In $\cal C$ there are two trivial regular solutions,
namely the vacua $f_0=0$
and $\bar{f}_0=\pi$
for which the energy (3) attains the global minimum $E_0=0$.

To my knowledge the only nonconstant regular solution of Eq.(4) known in
literature is an identity map $f=\psi$ of degree one (and the
corresponding antipodal map $\bar{f}=\pi-\psi$ of degree minus one).

In this paper I shall prove that Eq.(4)
has a countable family of regular solutions of degree zero and one,
which will be referred to as even and odd solutions respectively.
For even solutions
\beq
f(\psi) = f(\pi-\psi) \, ,
\eeq
while for odd solutions
\beq
f(\psi) = \pi - f(\pi-\psi) \, .
\eeq
\section{Numerical results}
In order to analyse  solutions of Eq.(4) it is convenient to introduce
a new independent variable $x=\ln(\tan(\psi/2))$ which changes from
$-\infty$ (for $\psi=0$) to $+\infty$ (for $\psi=\pi$), and define $h=f-\pi/2$.
In terms of these variables Eq.(4) reads
\beq
h'' -\tanh(x) h' + \sin(2h) = 0 \, .
\eeq
Under the reflection $x \ra -x$ even solutions are symmetric $h(x)=h(-x)$,
 while
odd solutions are antisymmetric $h(x)=-h(-x)$, hence it
 is sufficient to
consider Eq.(8) for $x \geq 0$. The formal power series
expansion near $x=0$ for odd solutions is
\beq
h(x) = b\, x + O(x^3) \, ,
\eeq
and for even solutions
\beq
h(x) = d + O(x^2) \, .
\eeq
 Since $x=0$ is a regular point of Eq.(8), these power
series have nonzero radii of convergence and therefore define local solutions
near $x=0$.

For regular solutions the asymptotic behaviour near $x=\infty$ is
\beq
\pm h(x) = -\frac{\pi}{2} + c \, e^{-x} + O(e^{-3x}) \, .
\eeq
I have integrated Eq.(8) numerically using a standard shooting procedure
for solving two-point boundary value problems [3]. The idea of this method
is to find such initial data (9) or (10) which give rise to correct asymptotic
behaviour (11).
 Below I describe the results for odd solutions, i.e., the initial
condition (9) (the behaviour of even solutions is analogous). For $b>1$
the function $h(x)$ increases monotonically, crosses $h=\pi/2$ at some $x_1$
and tends to $+\infty$ (as $e^{2x}$) for $x \ra \infty$. As $b \ra b_1=1$,
the point $x_1$ moves to $+\infty$ and for $b=1$ one gets the first odd
solution $h_1$. Actually this solution is nothing else but the identity map
 $h_1=\pi/2-2\arctan(e^x)$. This is the only solution which is known
analytically.
For $b$ slightly below one the function $h(x)$ after reaching a maximum
 monotonically decreases to $-\infty$, crossing $h=-\pi/2$
at some $x_3$.
 As $b \ra b_3= 0.29696...$ the point $x_3$
moves to $+\infty$. The value $b_3$ corresponds to the second odd solution
$h_3$.
 As $b$ decreases to zero this behaviour repeats
itself, i.e., there is a countable
sequence of initial values $b_n>0$ ($n$ odd) for which the solution $h_n(x)$,
staying within the strip $(-\pi/2,\pi/2)$, oscillates $\frac{(n-1)}{2}$
times around $h=0$  and goes to $h_n(\infty)=(-1)^{\frac{n-1}{2}} \pi/2$.

Similarly, there is a countable set of initial values $d_n$ defining
even solutions.
Both for odd and even solutions the index $n$ is equal to the number of
zeros of $h_n(x)$.

The numerical results are displayed in Tables~1,~2,~3,~4 and Figs~1,~2.
In Tables~3 and ~4 the symbol $x_k^n$ denotes the location of $k$th
nonnegative node of $h_n(x)$. For aesthetic reasons,
in Figs~1 and~2, I used the symmetry $h(x) \ra -h(x)$ to have the same
asymptotics for all solutions.
\begin{table} [h]
\caption{Parameters of odd solutions}
$$
\begin{tabular}{|c|c|c|c|c|c|} \hline
$n$ & $b_n$ & $c_n$ & $b_n^2/b_{n+2}^2$ & $c_{n+2}/c_n$ & $E_n/(12\pi^2)$ \\
 \hline
1 & 1  & 2 & 11.3397 & 12.1254 & 0.5\\
3 & 0.29696052 & 24.25080426 & 10.7816 & 10.8061 & 0.65168569\\
5 & 0.09043935 & 262.0555747 & 10.7519 & 10.7538 & 0.66527636 \\
7 & 0.02758136 & 2818.098290& 10.7493 & 10.7495 & 0.66653735 \\
9 & 0.00841249 & 30293.20888 & 10.7491 & 10.7491 & 0.66665464 \\
\hline
\end{tabular}
$$
\end{table}
\begin{table}
\caption{Parameters of even solutions}
$$
\begin{tabular}{|c|c|c|c|c|c|} \hline
$n$ & $d_n$ & $c_n$ & $d_n^2/d_{n+2}^2$ & $c_{n+2}/c_n$ & $E_n/(12\pi^2)$ \\
 \hline
2 & 0.33480103 & 7.07540708 & 11.1126 & 11.0168 & 0.61595784 \\
4 & 0.10043365 & 77.9483657 & 10.7794 & 10.7676 & 0.66199596 \\
6 & 0.03059018 & 839.318687 & 10.7519 & 10.7508 & 0.66623251 \\
8 & 0.00932911 & 9023.30712 & 10.7493 & 10.7492 & 0.66662628 \\
10 & 0.00284544 & 96993.7060 & 10.7491 & 10.7491 & 0.66666291 \\ \hline
\end{tabular}
$$
\end{table}
\begin{table}
\caption{Location of zeros for odd solutions}
$$
\begin{tabular}{|c|c|c|c|c|c|} \hline
$n$ & $x_1^n$ & $x_2^n$ & $x_3^n$ &  $x_4^n$ &  $x_5^n$   \\ \hline
1 & 0 & & & &  \\
3 & 0 & 2.230560  & & & \\
5 & 0 & 2.194022 & 4.606835  & & \\
7 & 0 & 2.190795 & 4.565917 & 6.982062  & \\
9 & 0 & 2.190496 & 4.562291 & 6.940749 & 9.356924  \\
$\infty$ & 0 & 2.190465 & 4.561920 & 6.936712 & 9.311532
 \\ \hline
\end{tabular}
$$
\end{table}
\begin{table}
\caption{Location of zeros for even solutions}
$$
\begin{tabular}{|c|c|c|c|c|c|} \hline
$n$ & $x_1^n$ & $x_2^n$ & $x_3^n$ &  $x_4^n$ &  $x_5^n$   \\ \hline
2 & 1.035870 & & & &  \\
4 & 1.012521 & 3.394669  & & & \\
6 & 1.010474 & 3.354995 & 5.770839  & & \\
8 & 1.010284 & 3.351483 & 5.729639 & 8.145811  & \\
10 & 1.010267 & 3.351158 & 5.725986 & 8.104472 & 10.520646 \\
$\infty$ & 1.010265 & 3.351124 & 5.725613 & 8.100431 & 10.475252 \\ \hline
\end{tabular}
$$
\end{table}
\section{Large {\em\bf n} behaviour}
For large $n$ the numerical data
exhibit  a striking scaling behaviour and the solutions seem to converge
 to some
limiting configuration. In order to understand this remarkable feature it is
 helpful to distinguish three characteristic regions.

In the inner region (I), $0 \leq x \, \lets \, 1$, $h_n(x)$ is small and
therefore
 is well
approximated by the solution of the linear equation
\beq
h'' - \tanh(x) h' + 2h = 0 \, .
\eeq
In this region we have approximately
\beq
h_n(x) = b_n \, g_i(x) \quad \mbox{for $n$ odd} \quad \mbox{and} \quad
h_n(x) = d_n \, g_i(x) \quad \mbox{for $n$ even} \, ,
\eeq
where the functions $g_i(x)$ ($i=1$ for odd and $i=2$ for even solutions)
solve Eq.(12) with initial conditions ($g_1(0)=0$,
$g_1'(0)=1$) and ($g_2(0)=1$, $g_2'(0)=0$), respectively.
As $n$ goes to infinity the zeros of $h_n(x)$ tend to the zeros of the
corresponding function $g_i(x)$, as is shown in Tables~3 and~4.

In the intermediate region (II), $1 \, \lets x \, \lets \, n$,
$h_n(x)$ is  still small (so (13) still holds) and
is well approximated by the solution of the linear equation
\beq
h'' - h' + 2h= 0 \, ,
\eeq
hence in this region the solution is almost periodic
\beq
h_n(x) \sim A_n e^{\frac{x}{2}} \sin (\frac{\sqrt{7}}{2} x + \delta_i ) \, ,
\eeq
where the amplitude $A_n$ is proportional to $b_n$ (or $d_n$)
 and the phase $\delta_i$ depends only on whether the solution is odd or even.

Finally, in the asymptotic region (III), $x \, \grts \, n$, $h_n(x)$ is no
longer
 small but can be well approximated by the solution of
\beq
h'' - h' + \sin(2h) =0 \, .
\eeq
The behaviour of this autonomous system determines the asymptotic behaviour
of solutions of Eq.(8).
Note that Eqs.(14) and (16) are translation invariant, i.e., if $h(x)$ is a
solution
so is $h(x+\Delta)$. Since asymptotically
$\pm h_n(x)= -\pi/2 + c_n e^{-x}$, we deduce that in regions II and III
approximately
\beq
\pm h_n(x) = s(x-\ln(c_n)) \, ,
\eeq
where $s(x)$ is the solution of Eq.(16) normalized by $s(x)=-\pi/2 + e^{-x}$
for $x \ra \infty$.

Now, consider two solutions $h_{n+2}$ and $h_n$. In regions II and III one has
from (17)
\beq
h_{n+2}(x) \simeq - h_n(x - \ln(\frac{c_{n+2}}{c_n})) \, .
\eeq
Since in region II the solution $h_{n+2}(x)$ has one zero more than
$h_n(x)$, and
 by Eq.(15)
the distance between two adjacent zeros is equal to $2\pi/\sqrt{7}$, one
obtains from (18) the relation
\beq
\ln \left(\frac{c_{n+2}}{c_n}\right) \simeq \frac{2\pi}{\sqrt{7}} \, ,
\eeq
which explains the numerical fact (see Table~1 and~2) that
\beq
\lim_{n \ra \infty}\frac{c_{n+2}}{c_n} = e^{\frac{2\pi}{\sqrt{7}}} \approx
10.7491 \, .
\eeq
The scaling relation for the initial parameters $b_n$ and $d_n$ may be deduced
in a
similar manner. Namely, it follows from (13) that in regions I and II one has
approximately
\beq
h_{n+2}(x) = \frac{b_{n+2}}{b_n}\, h_n(x) \quad  \mbox{for $n$ odd}
\quad \mbox{and} \quad h_{n+2}(x) = \frac{d_{n+2}}{d_n}\, h_n(x) \quad
\mbox{for $n$ even} \,.
\eeq
Combining (21) with (15), (18) and (19) one obtains
\beq
\frac{d_n^2}{d_{n+2}^2} \approx \frac{b_n^2}{b_{n+2}^2} \approx
 \frac{c_{n+2}}{c_n} \approx e^{\frac{2\pi}{\sqrt{7}}} \, .
\eeq
{}From this it follows in turn that the amplitude of oscillations of $h_n(x)$
in region II (and I) behaves as $A_n \sim e^{-\frac{n\pi}{\sqrt{7}}}$. Thus,
as $n \ra \infty$, the solution $h_n(x)$ tends pointwise to zero for any finite
$x$. The limiting solution $h_{\infty}=0$ is singular (because it fails to
satisfy the boundary condition (11) and therefore its energy
density $\rho$ diverges at north and south poles), nevertheless its energy is
finite $E_{\infty}=\frac{2}{3} \times 12\pi^2$.

Actually it is easy to show that $E=8 \pi^2$ is the upper bound for energy of
regular solutions. To see this rewrite Eq.(8) in the form
\beq
\left(\frac{1}{\cosh{x}} h' \right)' = - \frac{1}{\cosh{x}} \sin(2 h) \, .
\eeq
Multiplying this by $h$ and integrating by parts yields an identity
\beq
\int_{-\infty}^{\infty} \frac{dx}{\cosh{x}} h'^2 =
\int_{-\infty}^{\infty} \frac{dx}{\cosh{x}} h \sin(2 h) \,  ,
\eeq
which implies that on-shell
\beq
E = 4 \pi \int_{-\infty}^{\infty} \frac{dx}{\cosh{x}} (h \sin(2 h)
+2 \cos^2(h) )  \,.
\eeq
It will be shown below that for regular solutions $-\pi/2 < h < \pi/2$.
Within this range the function $F(h) \equiv h \sin(2 h) + 2 \cos^2(h)$ has
a maximum at $h=0$, hence the maximal energy is
\beq
E = 8 \pi \int_{-\infty}^{\infty} \frac{dx}{\cosh{x}} = 8 \pi^2 \, .
\eeq
Finally, using the scaling relations derived above one can also
 show that for large $n$
\beq
\frac{E_{\infty}-E_{n}}{E_{\infty}-E_{n+2}} \approx e^{\frac{2\pi}
{\sqrt{7}}} \, .
\eeq
Numerical fit gives the folowing empirical formula for energy
\beq
\frac{1}{12\pi^2} E_n = \frac{2}{3} - c\, e^{-\frac{n\pi}{\sqrt{7}}} \, ,
\eeq
where $c \approx 0.526$ for odd solutions and $c \approx 0.539$ for even
solutions.

\section{Stability analysis}

Now I turn to the stability analysis of harmonic maps described
above.
The Hessian of the energy functional at $h$ is
\beq
\delta^2 E(h) (\xi,\xi) = 4\pi \int_{-\infty}^{\infty} \left( \xi'^2
- 2 \cos(2h)\, \xi^2 \right)\, \frac{dx}{\cosh{x}} \, ,
\eeq
 which leads to an eigenvalue problem
\beq
-(\frac{1}{\cosh{x}} \, \xi')' - \frac{2}{\cosh{x}} \cos(2h)\, \xi =
 \frac{\lambda}{\cosh^3{x}}
\xi \,.
\eeq
Let  $\xi_k^n$ (and resp. $\lambda_k^n$) denote the $k$th
(where $k=1,2,...$)
eigenfunction (eigenvalue) around the regular solution $h_n$.
For  $h_1=\pi/2-2\arctan(e^x)$ Eq.(30) is solved by
\beq
\xi_k^1 = \frac{1}{\cosh{x}} \,C_{k-1}^2 (\psi)\, ,
 \qquad \lambda_k^1 = -4 + k (k+2) \, ,
\eeq
where $\psi=2\arctan(e^x)$ and
$C_k^2$ are the Gegenbauer polynomials [4]
 (for example $C_0^2=1$, $C_1^2=\cos\!\psi$).
Thus the solution $h_1$ has one unstable mode $\xi_1^1 = \frac{1}{\cosh{x}}$
with the eigenvalue $\lambda_1^1=-1$.

Actually all solutions $h_n$ (for $n \geq 1$) have an unstable mode
with the eigenvalue $-1$.
To see this consider a perturbation induced by the conformal Killing vector
field
on $S^3$  $K= \sin\! \psi \frac{\partial}{\partial \psi}$. In terms of
$x$-coordinate $K$
is the generator of translations $K=\frac{\partial}{\partial x}$
(in fact this is the defining property of the coordinate $x$), hence
\beq
\xi_{conf} \equiv \pounds_K h =  h'(x) \, .
\eeq
It is easy to check by differentiating Eq.(8) that $\xi_{conf}$
  satisfies
Eq.(30) with $\lambda_{conf}=-1$. Since by construction $\xi_{conf}$
has  $n-1$ nodes,
we conclude that $h_n$ has at least $n$ negative eigenvalues [5]. Numerics
shows that $\lambda_{conf}=-1$ is the greatest negative eigenvalue
for  each $h_n$
and therefore $h_n$ has
exactly $n$ unstable ($SO(3)$-symmetric) modes (see Table~5)
\footnote{It might seem from Table~5 that some eigenvalues are degenerate,
in fact
they form exponentially close pairs, which is a characteristic property
of a double-well potential with high barrier. I thank H.~Grosse and
T. Hoffman-Ostenhof for pointing this out to me.}.
\begin{table} [h]
\caption{Spectrum of perturbations}
$$
\begin{tabular}{|c|c|c|c|c|c|} \hline
$n$ & $\lambda_1^n$ & $\lambda_2^n$ & $\lambda_3^n$ & $\lambda_4^n$ &
$\lambda_5^n$ \\ \hline
1 & -1 & 4 & 11 & 20 & 31 \\
2 & -2.89259 & -1 & 4.59106 & 12.5133 & 22.7280 \\
3 & -10.6547 & -10.6108 & -1 & 4.41078 & 12.6498 \\
4 & -94.7578 & -94.7578 & -3.06867 & -1 & 4.62470 \\
5 & -1054.57 & -1054.57 & -10.6993 & -10.6506 & -1 \\ \hline
\end{tabular}
$$
\end{table}
This is a very interesting property which strongly suggests that the
existence of solutions $h_n$ is due to some topological Morse-theory
mechanism. This problem is currently under investigation.

It is of interest to study how the energy behaves under the action of the
conformal transformation. Given any solution $h(x)$ a conformally deformed
configuration $h^{\alpha}$ is given by
\beq
h^{\alpha}(x) = h(x+\alpha) \, .
\eeq
For the identity solution one has $h_1^{\alpha}=-\pi/2+2
 \arctan(e^{x+\alpha})$.
This one-parameter family of conformally deformed configuration defines
  a path in the function space which passess through $h_1$ for $\alpha=0$ and
 goes (non-uniformly) to the vacua
$h=\pm \pi/2$ for $\alpha \ra \pm \infty$. It is easy to compute the energy
along this path
\beq
E_1^{\lambda} = \frac{12 \pi^2}{1+\cosh{\alpha}} \, ,
\eeq
which, of course, for small $\alpha$ reproduces the linearized stability result
\beq
E_1^{\alpha} \simeq
E_1 - \alpha^2 \int_{-\infty}^{\infty} \frac{dx}{\cosh^3{x}}
(\xi_1^1)^2 = 6 \pi^2 - \frac{3}{2} \, \alpha^2  \, .
\eeq
Note that the path of conformally deformed configurations is
the path of steepest descent of energy from the saddle at $h_1$.

I have not studied
non-symmetric (i.e., depending on angles ($\vt,\vp$)) perturbations.
 However for the identity map $h_1$ it is known that
it has exactly four unstable degenerate eigenmodes with $\lambda=-1$, which
are induced by four linearly independent conformal Killing vector
fields on $S^3$ [6,7].

\section{Proof of existence}
The detailed proof of existence of harmonic maps described above will be
published separately.
In this section I shall only sketch the main idea of the proof  without going
into the technical details.

As I already mentioned, the asymptotic behaviour of solutions is determined
by the autonomous system
\beq
h'' - h' + \sin(2h) =0 \, .
\eeq
This equation has no regular solutions, that is, an orbit which runs into
one critical saddle-point, say $(h=-\pi/2,h'=0)$, when propagated
backwards, cannot reach another critical
saddle-point  $(h=\pi/2,h'=0)$ because it gets trapped by a focus at
$(h=0,h'=0)$ and spirals infinitely many times around it.
It is useful to regard the coefficient $\tanh(x)$ in front of $h'$
in Eq.(8) as a regularizing term which doesn't change the behaviour
of orbits of Eq.(36)
for large $x$ but regulates the behaviour at $x=0$ by changing the sign of
the
friction term at $x=0$. Actually the observation that for large $x$ Eq.(8)
is a perturbation of Eq.(36)
might be a basis of the existence proof (cf.[8]).  However this would require
making reference to some strong results on structural stability of phase
flow which I prefer not to invoke for a good reason that
 the problem may be tackled in much more straightforward manner.

I start with elementary a priori properties of regular solutions of Eq.(8).
\begin{lemma}
There exist two one-parameter families of local solutions of Eq.(8)
near $x=0$ analytic in $x,b,d$ such that
\beq
h(x) = bx + O(x^3) \, ,
\eeq
\beq
h(x)= d + O(x^2) \, ,
\eeq
where (37) and (38) correspond to odd and even solutions respectively.
\end{lemma}
{\em Proof.} Since $x=0$ is a regular point of Eq.(8), this is a
standard textbook theorem. Hereafter the solutions starting with initial
data (37) or (38) will be called $b$-orbits or $d$-orbits, respectively.
\begin{lemma}
There exits a one-parameter family of local solutions of Eq.(8) near
$x=\infty$ analytic in $c$ and $1/x$ such that
\beq
\pm h(x) = -\frac{\pi}{2} + c \, e^{-x} + O(e^{-3x}) \, .
\eeq
\end{lemma}
{\em Proof.} This fact is nontrivial since $x=\infty$ is a singular
point of Eq.(8). It may be proven using the desingularization technique
described in Proposition 1 in [8]. A $b$ or $d$-orbit will be called a
connecting orbit if it is regular for all $x \geq 0$ and satisfies (39)
for $x \ra \infty$.
\begin{lemma}
$b$-orbits and $d$-orbits exist for
 all $x \geq 0$.
\end{lemma}
{\em Proof.}
Integrating Eq.(23) from zero to some $x>0$ yields
\beq
\frac{1}{\cosh{x}} h'(x) = h'(0) - \int_0^x \frac{dx}{\cosh{x}} \sin(2h)
\, .
\eeq
Thus $h'(x)$ is finite for any $x \geq 0$ and thereby $h(x)$ is also finite.
\begin{lemma}
If $h(x)$ leaves the strip $(-\pi/2,\pi/2)$, then it monotonically tends to
$\pm \infty$.
\end{lemma}
{\em Proof.} Define a function
\beq
W(h,h') = \frac{1}{2} h'^2 + \sin^2{h} \, .
\eeq
Using Eq.(8) one has
\beq
\frac{dW}{dx} = \tanh{x}\, h'^2  \, ,
\eeq
hence $W$ increases with $x>0$. Suppose that
$h(x_0)=\pi/2$ and $h'(x_0)=a>0$ for some $x_0 >0$. Then
 for $x>x_0$
$$
W(x)-W(x_0) = \frac{1}{2} h'^2(x) + \sin^2{h(x)} -\frac{1}{2} a^2 -1 >0
\, ,
$$
which implies that $h'^2(x)>a^2$.
\begin{cor}
There are no regular solutions of Eq.(8) with homotopy degree bigger than one.
\end{cor}
\begin{lemma}
If $h(x)\in (-\pi/2,\pi/2)$ for all $x>0$, and has a finite number of zeros,
then $h' \ra 0$ and $h \ra \pm \pi/2$ as $x \ra \infty$.
\end{lemma}
{\em Proof.} First note that within the strip $(-\pi/2,\pi/2)$ $h(x)$ has no
positive minima nor negative maxima, as follows immediately from Eq.(8).
Thus
 for sufficiently large $x$ the solution $h(x)$ is monotonic
 and therefore has a limit
as $x \ra \infty$. This implies that $h'(\infty)=0$, hence by Eq.(8)
$h(\infty)=\pm \pi/2$ or $0$. To complete the proof one has to show that the
case $h(\infty)=0$ is impossible. To prove this
suppose that $h(\infty)=0$ and $x_0$ is the last extremum of $h$. Without loss
of generality one may assume that $h$ has a maximum at $x_0$, hence
$h(x) \geq 0$ for all $x \geq x_0$. Now, multiply Eq.(23) by $\cos{h}$
and integrate by parts from $x_0$ to infinity. This gives
\beq
\int_{x_0}^{\infty} \frac{dx}{\cosh{x}} \, \sin{h} \, ( h'^2 + 2 \cos^2{h} )
= 0 \, ,
\eeq
which is a contradiction unless $h\equiv 0$.

{\em Remark.} In fact Lemma~5 is true without assuming that $h$ has a finite
number of zeros, however then the proof is  more complicated (and this
stronger result is not needed for the proof of existence).

Now, having established the elementary properties of solutions, one
 can proceed
with the proof. The main idea is taken from
 the proof of existence of
globally reglar and black hole solutions of Einstein-Yang-Mills
 equations given recently by Smoller,  Wasserman and Yau [9,10].

For concreteness  I shall outline  the proof in the case
of odd solutions (the proof of
existence of even solutions is completely analogous). After [9] I introduce
the following notation.  Define the region $\Gamma$ by
\beq
\Gamma = \{ (h,h',x) : |h| < \frac{\pi}{2} , x>0, (h,h') \neq (0,0) \}
\, .
\eeq
Next, let $x_e(b)$ be the smallest $x>0$ at which the $b$-orbit exits $\Gamma$.
For any $b$-orbit define $\theta(x,b)$ by
\beq
\theta(0,b) = -\frac{\pi}{2} \quad \mbox{and} \quad
\theta(x,b) = \arctan \left(\frac{h'(x,b)}{h(x,b)} \right) \quad \mbox{for}
\, x>0 \, .
\eeq
The rotation number, $\Omega(b)$, of the $b$-orbit is given by
\beq
\Omega(b) = -\frac{1}{\pi} \left( \theta(x_e(b),b)+\frac{\pi}{2} \right) \, .
\eeq
Note that for connecting orbits $\Omega=n/2$ with $n$ odd.

\begin{theorem}
For each odd $n$ there exists a connecting $b_n$-orbit with
$\Omega(b_n)=n/2$.
\end{theorem}
The proof is based on three technical propositions.
\begin{prp}
If $b>\sqrt{2}$, the $b$-orbit must exit $\Gamma$ through $h=\pi/2$ with
$\Omega(b)<1/2$.
\end{prp}
{\em Proof.} For $b>\sqrt{2}$, $W(0)>1$ and therefore $W(x)>1$ for $x>0$
which implies that $h'(x)$ is positive for all $x>0$.

The following two technical propositions are crucial for the argument. They are
analogues  of Proposition 3.5 in [10] and Proposition 3.4 in [9],
and their proofs follow closely, with gross simplifications, the proofs
of these  propositions.
\begin{prp}
Given any $N>0$, there is an $\epsilon>0$, such that if $0<b<\epsilon$,
then $\Omega(b)>N$.
\end{prp}
This is a compactness property which guarantees that $b$-orbits with bounded
rotation have $b$ greater than some positive constant.
\begin{prp}
Suppose that a $b_n$-orbit is a connecting orbit with $\Omega(b_n)=n/2$. Then
 for
sufficiently small $\epsilon$, for $b_n-\epsilon<b<b_n$, a $b$-orbit exits
$\Gamma$ transversally through $|h|=\pi/2$ with $n/2<\Omega(b)<1+n/2$.
\end{prp}
This says that orbits which are close to connecting orbits (with $b<b_n$)
exit
$\Gamma$ without making another rotation.

Now, we are ready to complete the proof of Theorem 1.
 The main idea is to construct connecting orbits step by step.

Step 1: Define
\beq
b_1 = inf \, \{b : \mbox{$b$-orbit exits $\Gamma$ via}\;
h=\frac{\pi}{2} \; \mbox{with}
\; \Omega(b) \leq \frac{1}{2} \} \, .
\eeq
Proposition 2 guarantees that $b_1>0$. Now, the $b_1$-orbit cannot exit
$\Gamma$
via $h=\pi/2$ because the same would be true for nearby orbits with $b<b_1$,
violating the definition of $b_1$. Thus the $b_1$-orbit stays in $\Gamma$ for
all $x>0$, hence by Lemma 5   it is a connecting orbit and $\Omega(b_1)=1/2$.

Step 2: Let
\beq
\bar b_1 = min \, \{ b_1: \mbox{$b_1$-orbit is a connecting orbit with} \;
\Omega(b_1)=\frac{1}{2}
\} \, .
\eeq
{\em Remark.} This step is necessary because I was not able to show
the uniqueness of the $b_1$-orbit.

Take $b$ slightly smaller than $\bar b_1$. By Proposition 3
 the $b$-orbit exits $\Gamma$ through
$h=-\pi/2$ with $1/2<\Omega(b)<3/2$. Define
\beq
b_3 = inf \, \{ b: \mbox{$b$-orbit exits} \; \Gamma \; \mbox{via} \;
h=-\frac{\pi}{2} \; \mbox{with} \;
\Omega(b) \leq \frac{3}{2} \} \, .
\eeq
By the completely analogous argument as above one concludes that $b_3$ defines
a connecting orbit with $\Omega(b_3)=3/2$. The subsequent connecting orbits
are obtained by the repetition of the above construction.

\section{Acknowledgments}
I am grateful to P. Aichelburg, R. Beig and T. Chmaj  for
discussions and comments.  Special thanks are due to J. Smoller for his
patience in explaining to me the details of his work. I also thank
 F. Burstall, J. Eells, L. Lemaire,
and J. Rawnsley for providing me with the information about literature
on harmonic maps, and especially for convincing me that my harmonic maps
were not known to mathematicians. This research was supported in part by the
Fundacion Federico and the KBN grant 2/P302/113/06.

\section*{Figure captions}
\begin{description}
\item[Fig.1] Odd solutions for $n=1,3,5,7,9$.
\end{description}
\begin{description}
\item[Fig.1] Even solutions for $n=2,4,6,8,10$.
\end{description}


\newpage
\begin{thebibliography}{17}
\bibitem{1} J. Eells and L. Lemaire, {\em A report on harmonic mappings} ,
Bull. London Math. Soc. {\bf 10} (1978) 1;\\
{\em Another report on harmonic mappings}, Bull. London Math. Soc.
{\bf 20} (1988) 385.
\bibitem{2} R.S. Palais, {\em The principle of symmetric criticality},
Commun. Math. Phys. {\bf 69} (1979) 19.
\bibitem{3} W.H. Press {\em et al.}, {\em Numerical Recipies},
(Cambridge University Press, New York, 1992).
\bibitem{4} M. Abramowitz and I.A. Stegun, {\em Handbook of mathematical
functions}, (Dover, New York, 1965) Chapter 22.
\bibitem{5} M. Reed and B. Simon, {\em Methods of Modern Mathematical
Physics, vol.IV},  (Academic Press, New York, 1978).
\bibitem{6} R.T. Smith, {\em The second variation formula for harmonic
mappings}, Proc. Amer. Math. Soc. {\bf 47} (1975) 229.
\bibitem{7} Y.L. Xin, {\em Some results on harmonic maps},
 Duke Math. J. {\bf 47} (1980) 609.
\bibitem{8} P. Breitenlohner, P. Forgacs and D. Maison, {\em On static
spherically symmetric solutions of the Einstein-Yang-Mills equations},
Commun. Math. Phys. {\bf 163} (1994) 141.
\bibitem{9} J.A. Smoller and A. Wasserman,
{\em Existence of infinitely many smooth, static, global solutions
of the Einstein-Yang/Mills equations}, Commun. Math. Phys. {\bf 151}
(1993) 303.
\bibitem{10} J.A. Smoller, A. Wasserman and S.-T. Yau,
{\em Existence of black hole solutions for the Einstein-Yang/Mills
equations}, Commun. Math. Phys.
 {\bf 154} (1993) 377.
\end{thebibliography}
\end{document}